\documentclass[a4paper]{article}
\usepackage{graphicx}
\usepackage{amsmath}

\begin{document}

\begin{center}
{\bf \Large
Remanence and switching sensitivity in nanodot magnetic arrays
}\\[5mm]

{\large
Przemys{\l}aw Gawro{\'n}ski and Krzysztof Ku{\l}akowski
}\\[3mm]

{\em
Department of Applied Computer Science,
Faculty of Physics and Applied Computer Science,
AGH University of Science and Technology\\
al. Mickiewicza 30, PL-30059 Krak\'ow, Poland
}

\bigskip
{\tt kulakowski@novell.ftj.agh.edu.pl}

\bigskip
\today
\end{center}

\begin{abstract}
New results are reported of the computer simulations on the magnetic behaviour of
magnetic arrays of nanoscopic dots, placed in cells of the square lattice. We show that the 
remanence magnetization $M_r$ decreases 
with the array size. For arrays 50x50, we investigate also the stability of the magnetic structure 
of an array in an oscillating magnetic field. The damage spreading technique reveals that 
this stability increases with the standard deviation $\sigma$ of the switching field of individual 
elements of the array. On the other hand, $M_r$ decreases with $\sigma$. An optimalization
of the system (large $M_r$ and large stability) can then be reached at some intermediate 
value of $\sigma$. 
\end{abstract}

\noindent
{\em PACS numbers:} 75.75.+a ; 75.40.Mg

\noindent
{\em Keywords:} hysteresis, computer simulation, magnetostatic interaction, switching field distribution

\section{Introduction}
Magnetic nanodot arrays have been proposed for use in magnetic storage media \cite{chou,ros1,brm}.
It is worthwhile, then, to study magnetic properties of these systems. In particular, the long-range
character of the magnetostatic interaction is expected to introduce correlations between magnetic 
states of neighboring dots. These correlations can deteriorate the quality of data storage. 
In our recent work \cite{kac1,kac2} we discussed the system of nickel dots, designed by 
the group of C. A. Ross in MIT \cite{ros2}. The aspect ratio of the dots is in this case large enough 
to assure the one-domain Ising character of the magnetic moments, which are directed up or down. 
We proved in
particular, that a reduction of the switching field fluctuations from one dot to another leads to an
instability of the magnetic structure of the array in the presence of a weak oscillating field. 
In this way, also the quality of the information stored there is lowered. 

Our aim is to report results of a further continuation of this research. Our subject here is a finite
array of magnetic dots; on the contrary to our previous approach we do not apply the periodic 
boundary conditions. The goal of this paper is {\it i)} the finite size effect displayed by the 
remanence magnetization, {\it ii)} further elucidation of the role of the standard deviation $\sigma$ of 
the switching field. We demonstrate, that the condition of large remanence and the condition of 
the stability are in conflict. As the standard deviation $\sigma$ increases, the stability 
is improved, 
but the remanence decreases. The latter means that at zero field, states close to saturation are
not possible. The magnetostatic interaction between dots removes these states from the accessible set,
and therefore they cannot be used to store an information. 

The text is organized as follows: the system parameters and our model assumptions are given in 
Section 2, numerical results - in Section 3. Last section is devoted to conclusions.

\section{From the system to the model}

Squared or rectangular arrays are formed from cylindrical dots of average diameter 57 $nm$,
average height 115 $nm$, with period 100 $nm$. The cylinders are perpendicular to the array. 
The material is crystalline nickel, and each dot is one 
domain. Saturation magnetization of the dots is $M_s$=370 $emu/cm^3$, the average switching field
is 710 $Oe$ and its standard deviation $\sigma$ is 105 $Oe$ \cite{ros2}. In our calculations, this 
deviation plays the role of model parameter, which can be larger or smaller, depending on how the 
technique of the array designing is refined. The probability distribution of the 
switching field is supposed to be Gaussian. 

In our model calculations, the magnetic moments of the dots are supposed to be Ising-like, i.e. up or
down. The dot-dot magnetostatic interaction is calculated within the Rectangular Prism Approximation 
(RPA) \cite{sch}. This formula has been checked to reproduce properly at least the order of magnitude of the 
interaction field \cite{ros3}. We should add that the problem how to model this 
interaction is far from trivial; this point is discussed thoroughly in \cite{vaz}. The system 
dynamics is reproduced within the Pardavi-Horvath algorithm \cite{par}. This algorithm, numerically 
efficient, properly reproduces the time order of switching of particular magnetic moments.

Recently, the repeatability of the switching behaviour of a magnetic state of a macroscopic ring
has been investigated in the presence of thermal fluctuations \cite{ros4}. In the simulation, a study 
of the repeatability can be performed by means of the damage spreading technique \cite{jan,kac1,kac2}. This 
method seems to be directly designed to detect
events when "changes in the configurations of small regions of spins are (...) likely to to be able to 
affect the macroscopic switching" \cite{ros4}. Having this in mind, we apply the damage spreading technique
to investigate the thermal stability of the system for various values of $\sigma$.

The algorithm of this technique is as follows. The configuration of magnetic moments (say, spins) 
of dots is chosen randomly 
and evolves to a stable state. Then, the state is modified by a change of one spin. Then we introduce
an oscillating magnetic field to the simulation. A spin flips if it is opposite to the effective (external 
plus interaction) field, and this field excesses its switching field. Both copies, the original and the modified
state, evolve simultaneously. The damage is the Hamming distance between the copies, i.e. the number of spins
of orientations different in the original and the modified states. Sometimes the damage becomes zero, sometimes
it evolves close to a constant or grows up to the range of the system size.

\section{Numerical results}

In Fig. 1 we show how the system size influences the remanence magnetization. For a system of only a few dots, 
the remanence is just equal to $M_s$. As the number of dots increases, we observe a clear decrease of $M_r$. 
The effect is less pronounced when $M_s$ is smaller; this is due to the interaction energy, which is 
proportional to $M_s^2$. The calculations are made also for rectangular arrays, with the side ratio 3:2. 
As we see, the plots for the square and rectangular arrays coincide.

\begin{figure}
\begin{center}
\includegraphics[angle=-90,width=.8\textwidth]{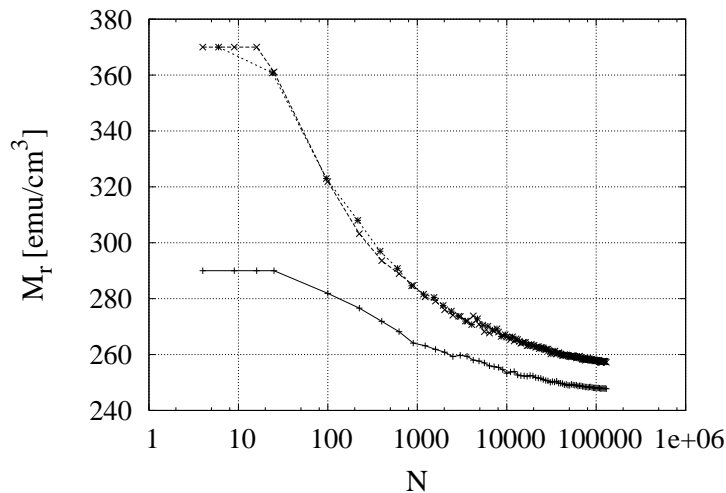}
\caption{The remanence magnetization $M_r$ against the number of dots $N$. 
The lines are for guide eyes. The lowest curve (+) is for square arrays, $M_s=290$ $emu/cm^3$. 
Two higher curves are for squared (x) and rectangular (*) arrays, $M_s=370$ $emu/cm^3$. 
The ratio of rectangle sides is 2:3.}
\label{Mr vs area}
\end{center}
\end{figure}

\begin{figure}
\begin{center}
\includegraphics[angle=-90,width=.8\textwidth]{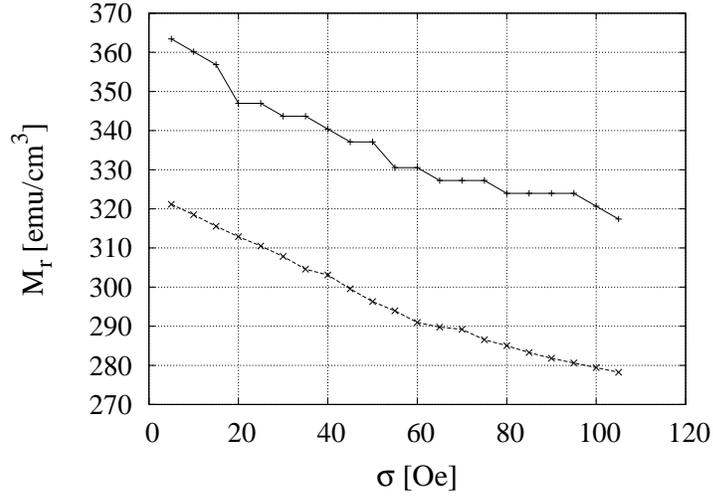}
\caption{The remanence magnetization $M_r$ against the standard deviation $\sigma$ of the switching field for two
arrays: $15\times 15$ (upper curve) and $50 \times 50$ (lower curve).}
\label{Mr vs sigma}
\end{center}
\end{figure}

The remanence magnetization is also calculated against $\sigma$, where the latter varies from 5 to 105 $Oe$. 
These results are shown in Fig. 2 for two arrays: $15\times 15$ and $50\times 50$. In accordance to Fig. 1,
the remanence is smaller for larger systems. As we can see in Fig. 2, $M_r$ decreases also with $\sigma$. The explanation
of this effect is as follows:
when the distribution of the switching field is broadened, the switching field of more and more dots become
smaller than the interaction field; then more and more spins flip. As a consequence, less and less states of the 
array are stable at zero field. Clearly, these unstable states cannot be used for information storage \cite{kac3}.
Then, an increase of $\sigma$ makes an array less useful to this purpose.

\begin{figure}
\begin{center}
\includegraphics[angle=-90,width=.8\textwidth]{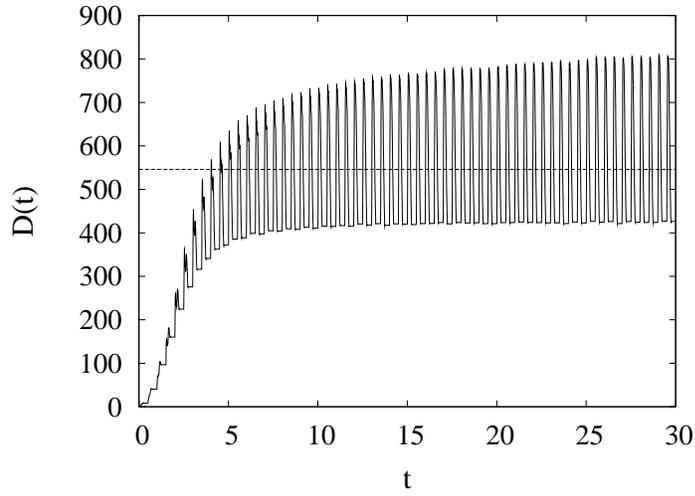}
\caption{The damage $D$ against numerical time $t$ for $\sigma=15$ $Oe$. The curve is an average over 200 cases, 
where the damage does not die out. Other 100 cases is not taken into account. The horizontal curve is the time average 
$<D>$ over one period of the applied field in the stationary state.}
\label{D(t) vs t}
\end{center}
\end{figure}

An example of the plot obtained by means of the damage spreading technique is shown in Fig. 3. The plot is an average $D(t)$
of the damage against time $t$ over the realizations, when the damage remains finite. For each realization, different random values 
of the switching field are assigned to 
all dots. The horizontal line marks the time average of the damage at a stable state, where the mean and the amplitude
of the oscillations remain constant. This average damage $<D>$ is shown in Fig. 4, as dependent on the standard deviation 
$\sigma$. As we see, the damage decreases with $\sigma$. This means, that once the switching fields of the dots differ
more, the magnetic state of the array is more stable. We should add that in some cases, the damage does not spread; instead
it vanishes immediately after the simulation starts. The probability of such vanishing increases with $\sigma$ 
from 33 percent for $\sigma$ = 15 $Oe$ to 63 percent for $\sigma$ = 105 $Oe$; in this way, the effect shown
in Fig.4 is even more strengthened. 

\begin{figure}
\begin{center}
\includegraphics[angle=-90,width=.8\textwidth]{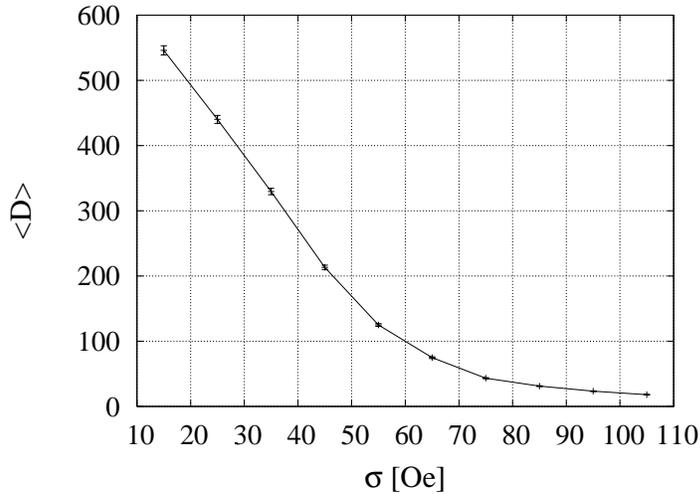}
\caption{The average damage $<D>$ against $\sigma$ for the array $50\times 50$. }
\label{D(t) vs sigma}
\end{center}
\end{figure}

\section{Conclusions}

Two kinds of stability, investigated here, reveal the opposite behaviour with the variation of the standard deviation
$\sigma$ of the switching field distribution. If $\sigma$ is small, the remanence $M_r$ is close to the saturation 
value $M_s$. This means, that the saturated magnetic state is stable or almost stable, as in most cases the switching field 
is larger than the interaction field. Then, all magnetic states can be used as different words 
in the message encoded in the system. However, in the same case of small $\sigma$ a small change of the magnetic
configuration spreads easily over the lattice in the presence of an oscillating magnetic field. In other words, this 
transport of spin flips appears at almost the same barrier of energy at neighoring dots \cite{ros4}. Here, the energy 
barrier is due to the switching field. 

On the contrary, for large $\sigma$ the damage does not spread, but it remains localized. This effect is analogous to 
the Anderson localization of electrons at disordered systems \cite{and}. This is desirable for the stability of the 
information stored; however, simultaneously the remanence is small, as some spins flip spontaneously under the interaction
field. 

Conluding, the stability of the magnetic configurations of the array are limited by two different mechanisms,
one prevailing for small and another for large value of the standard deviation $\sigma$ of the switching field. 
Apart of technological 
difficulties, an optimal value of $\sigma$ does depend on a particular application.

\bigskip
{\bf Acknowledgements}. The authors are grateful to Prof. Julian Gonz{\'a}lez for helpful 
discussions and kind hospitality.

\bigskip

\end{document}